\documentclass[reprint,amsmath,amssymb,aps]{revtex4-1}
\usepackage{graphicx}
\usepackage{dcolumn}
\usepackage{bm}
\usepackage{color}
\usepackage[margin=2cm]{geometry}
\usepackage{epstopdf}

\begin{document}

\title{Structural Diversity in Lithium Carbides}
\author{Yangzheng Lin$^1$}
\author{Timothy A. Strobel$^1$} \email{tstrobel@carnegiescience.edu}
\author{R. E. Cohen$^{1,2}$} \email{rcohen@carnegiescience.edu}
\affiliation{$^1$Extreme Materials Initiative, Geophysical Laboratory, Carnegie Institution of Washington, 5251 Broad Branch Road, NW, Washington, DC 20015, USA}
\affiliation{$^2$University College London, UK}

\begin{abstract}
The lithium-carbon binary system possesses a broad range of chemical compounds, which exhibit fascinating chemical bonding characteristics that give rise diverse and technologically important properties. While lithium carbides with various compositions have been studied or suggested previously, the crystal structures of these compounds are far from well understood. In this work we present the first comprehensive survey of all ground state (GS) structures of lithium carbides over a broad range of thermodynamic conditions, using \textit{ab initio} density functional theory (DFT) crystal structure searching methods. Thorough searches were performed for 29 stoichiometries ranging from Li$_{12}$C to LiC$_{12}$ at 0 GPa as well as 40 GPa. Based on formation enthalpies from optimized van der Waals density functional calculations, three thermodynamically stable phases (Li$_4$C$_3$, Li$_2$C$_2$ and LiC$_{12}$) were identified at 0 GPa, and seven thermodynamically stable phases (Li$_8$C, Li$_6$C, Li$_4$C, Li$_8$C$_3$, Li$_2$C, Li$_3$C$_4$, and Li$_2$C$_3$) were predicted at 40 GPa. A rich diversity of carbon bonding, including monomers, dimers, trimers, nanoribbons, sheets and frameworks, was found within these structures, and the dimensionality of carbon connectivity existing within each phase was observed to increase with increasing carbon concentration. Of particular interest, we find that the well-known composition LiC$_6$ is actually a metastable one. We also find a unique coexistence of carbon monomers and dimers within the predicted thermodynamically stable phase Li$_8$C$_3$, and different widths of carbon nanoribbons coexist in a metastable phase of Li$_2$C$_2$ (\textit{Imm}2). Interesting mixed \textit{sp}$^2$-\textit{sp}$^3$ carbon frameworks are predicted in metastable phases with composition LiC$_6$.
\end{abstract}

\pacs{}
\keywords{}

\maketitle


\section{Introduction}
Numerous novel carbon allotropes \cite{Hoffmann83,Johnston89,Blase04,Itoh09,Pickard10,Wang10a,Zhao11,Zhu11,Lyakhov11,Hu12,Zhao12,Jiang13,Zhang13,Niu14,Bu14,Hu14} have been predicted theoretically over the past decades. Some of these proposed structures have excellent mechanical, optical and/or electronic properties, which are important for a wide range of potential applications. For example, three-dimensional (3D) metallic carbon allotropes \cite{Hoffmann83,Itoh09,Zhang13,Niu14,Bu14,Hu14} are potentially important conductors with excellent chemical inertness under ambient conditions, and carbon allotropes with high elastic constants but low densities like clathrates \cite{Timoshevskii02,Blase04,Pickard10,Wang10a} would be especially useful for light-weight structural materials. However, it is particularly challenging to synthesize these materials from pure carbon, due to their relatively higher enthalpies than graphite or diamond, and only limited experimental evidence for these phases currently exists \cite{WangY12,Amsler12}. Another way to synthesize pure carbon allotropes is to start from carbide precursors. This approach has been successful in the production of so-called carbide-derived carbon \cite{Presser11}. Novel silicon and germanium allotropes may be produced through the leaching of metal atoms from metal silicide or germanide precursors \cite{Guloy06,Connetable10,Kurakevych13b,Kim15}, which suggests the possibility of making pure carbon allotropes from metal carbides in a similar way. Considering that carbon atoms have a smaller radius than silicon atoms, we focus on the possibility of carbon framework-based lithium carbides. In order to establish potential thermodynamic stability for these types of structures, we have investigated Li-C compounds over a broad compositional range to understand which kinds of precursors might exist under ambient and high-pressure conditions, and to gain insights into the forms of carbon existing within them.

A series of lithium carbon binary compounds (including Li$_6$C\cite{Schleyer83,Reed85,Kudo92}, Li$_5$C\cite{Schleyer83,Reed85}, Li$_4$C\cite{Chung72,Shimp73,Collins76,Shimp81,Landro83,Schleyer83,Reed85,Ruschewitz99}, Li$_3$C\cite{Chung72,Morrison75,Sneddon75,Shimp79,Landro83,Ruschewitz99}, Li$_8$C$_3$\cite{Ruschewitz99}, Li$_2$C\cite{Morrison75,Sneddon75,Shimp81,Shimp79,Ruschewitz99}, Li$_4$C$_3$\cite{West65,West69,Shimp73,Morrison75,Sneddon75,Jemmis77,Shimp81,Marynick96,Ruschewitz99}, Li$_2$C$_2$\cite{Rober67,Shimp73,Dill79,Shimp79,Shimp81,Avdeev96,Ruschewitz99}, Li$_4$C$_5$\cite{Chwang73,Ruschewitz99}, LiC$_2$\cite{Mordkovich96}, LiC$_3$\cite{Mordkovich96}, LiC$_6$\cite{Guerard75,Holzwarth83,Avdeev96}, LiC$_{10}$\cite{Yazami01}, and LiC$_{12}$\cite{Guerard75,Holzwarth83,Avdeev96}, etc.) have been reported theoretically or experimentally since half a century ago. Most of those previous studies focused on the molecular structure and the stability of molecular clusters. Of these reported compounds, only  Li$_2$C$_2$\textcolor{red}{\cite{Rober67,Ruschewitz99,Ruprecht10,Chen10,Nylen12,Benson13,Drue13,Tang15,Efthimiopoulos15}} and some graphite intercalation compounds (LiC$_6$ and LiC$_{12}$, etc.)\cite{Wertheim80,Holzwarth83,Andersson99,Kganyago03,WangX12,Drue13,Hazrati14} have been investigated experimentally in their crystal structures.

Ab initio density functional theory calculations have many successful examples of predicting the relative stabilities for solid crystal phases of single elements and multicomponent compounds under ambient and high-pressure conditions. Some methods including random sampling \cite{Pickard06,Pickard11}, minima hopping \cite{Goedecker04,Amsler10} and those implemented in \textsc{Uspex} \cite{Oganov06,Lyakhov13}, \textsc{Calypso} \cite{Wang10b,Wang12} or \textsc{XtalOpt} \cite{Lonie11,Lonie12} were developed in the past decade and have made the prediction of ground state crystal structures much easier and more efficient \cite{Zurek15}. In the system of lithium carbon compounds, some searches were performed previously for the stoichiometry Li$_2$C$_2$ \cite{Chen10,Nylen12,Benson13} and some polymeric forms of carbon were predicted at high pressure \cite{Chen10,Benson13}. In this work, we predict two more stable high-pressure phases of Li$_2$C$_2$. We provide the static convex hulls (\textit{i.e.}, formation enthalpy vs concentration diagrams) at 0 and 40 GPa, and thus predict a broad range of novel, thermodynamically stable, lithium carbide phases. Various forms of carbon are found to exist within these stable crystal structures, and suggest energetically viable pathways to novel carbon /textcolor{red}{materials}.

\section{Computational Methods}

We predict the GS crystal structures of lithium carbides through evolutionary algorithm-based searching methods, as implemented in the opensource package \textsc{XtalOpt} \cite{Lonie11,Xtalopt14}. The evolutionary algorithms in \textsc{XtalOpt} were designed to generate new structures that have lower enthalpies than structures in previous generations. All searches were initialized by 30--60 randomized or specified structures, and they were not terminated until the lowest enthalpy structure survived after 300--600 generated structures. Each search used a fixed number of formula units of Li$_m$C$_n$ ($m$ and $n$ are irreducible integers except in Li$_2$C$_2$, where the reducible notation is preserved following standard convention). The structure with the lowest enthalpy is regarded as the ground state for a given stoichiometry. At a given pressure and stoichiometry, several searches with different numbers (1--6) of formula units were performed to avoid missing the true ground state structure. For computational efficiency, the largest number of primitive cell atoms was limited to 20 in all searches. The symmetries of low-enthalpy structures were refined using \textsc{FindSym} \cite{Stokes05,ISOtropy14}.

The enthalpy of each structure within the evolutionary algorithm searching was calculated from DFT relaxation using the projector-augmented wave (PAW) method \cite{Blochl94,Kresse99} within the Perdew-Burke-Ernzerhof (PBE) generalized gradient approximation (GGA) \cite{Perdew96,Perdew97}. The DFT structural relaxations were performed using \textsc{pwscf} in the package of \textsc{Quantum-Espresso} \cite{Giannozzi09,Pwscf14}. In our DFT calculations, the electronic configurations for Li and C were 1s$^2$2s$^1$ and [He]2s$^2$2p$^2$, respectively. The plane-wave kinetic-energy cutoff was 80 Ry (1088 eV). During the searching process, the Monkhorst-Pack (MP) k-point meshes $k_1\times k_2\times k_3$ were given according to $k_i=b_i/(2\pi\times 0.06)$ ($i=1,2,3$) where $b_1$, $b_2$ and $b_3$ are the lattice lengths (in unit of $\text{\AA}^{-1}$) in reciprocal space. The relative enthalpies of lithium carbides converged up to 6 meV/atom within these settings. For each stoichiometry, several low-enthalpy structures were selected carefully from all the crystal structures obtained by searchings. DFT relaxations were reinvestigated for these selected structures with denser k-point meshes of $k_i=b_i/(2\pi\times 0.04)$ (the lower limit of $k_i$ was 2). The relative enthalpies of lithium carbides converged up to 2 meV/atom with these denser k-point meshes. In the calculations of enthalpies vs pressures, the k-point meshes were fixed within one structure at different pressures. Even denser k-point meshes of $k_i=b_i/(2\pi\times 0.02)$ were used in our density of state (DOS) calculations for all the thermodynamically stable and some metastable structures. For all the thermodynamically stable and metastable structures, we calculated the phonon frequencies using density-functional perturbation theory (DFPT) to determine their dynamic stabilities.

Twenty-nine stoichiometries of Li$_m$C$_n$ ($m:n$ in 12:1, 10:1, 8:1, 6:1, 5:1, 4:1, 3:1, 8:3, 5:2, 2:1, 5:3, 3:2, 4:3, 5:4, 2:2, 4:5, 3:4, 2:3, 3:5, 1:2, 2:5, 3:8, 1:3, 1:4, 1:5, 1:6, 1:8, 1:10 and 1:12), which include all the possible ground state lithium carbides between Li$_{12}$C and LiC$_{12}$ suggested by previous studies (see the introduction), have been investigated in order to determine the GS structures. While we consider our searching to be comprehensive over a broad range of composition, we acknowledge the finite nature of these searches given limitations of computational resources, and realize the possibility of thermodynamically stable phases for unconsidered stoichiometries or number of primitive cell atoms. For a given pressure, the thermodynamically stable structures are those whose formation enthalpies per atom lie on the convex hull of formation enthalpy a function of composition \cite{Zeng13,Morris14}. For a compound Li$_m$C$_n$, the formation enthalpy of a structure under pressure $P$ is defined as

\begin{equation}
\Delta H(P)=\frac{H_{\text{Li}_m\text{C}_n}(P)-mH_{\text{Li}}(P)-nH_{\text{C}}(P)}{m+n}
\label{eq:DeltaH}
\end{equation}
where $H_{\text{Li}_m\text{C}_n}$ is the enthalpy per formula unit of Li$_m$C$_n$ for a given structure. $H_{\text{Li}}$ and $H_\text{C}$ are enthalpies per atom of lithium and carbon in their ground-state crystal structures, respectively. The atomic concentration of carbon in the compound Li$_m$C$_n$ is defined as,
\begin{equation}
x_{\text{C}}=\frac{n}{m+n}
\label{eq:ConC}
\end{equation}

\section{Results and Discussions}
\subsection{Thermodynamically Stable Lithium Carbides}

The van der Waals (vdW) interaction has proved to be important for predictions of both structural and energetic information for graphite and graphite intercalated lithium compounds \cite{Lee12,Hazrati14,Ganesh14}. Here we show in Table \ref{tab:FEvdW} that the vdW interaction is also crucial in prediction of the formation energies of Li$_2$C$_2$. Except in the DFT-local density approximation (DFT-LDA) calculations where the LDA PAW pseudopotentials (PAWs) were used, we used the PBE PAWs in all other DFT and vdW calculations. All the corresponding pseudopotentials of lithium and carbon were taken from the pseudopotential library of \textsc{Quantum-Espresso} \cite{Giannozzi09,Pwscf14}. In DFT-D2 \cite{Grimme06} calculations, the default atomic parameters were used without modifications. The vdW, vdW2, optB88-vdW, optB86-vdW, rev-vdW2 \cite{Dion04,Klimes10} calculations shared the same vdW kernal table, while the rVV10 \cite{Sabatini12} calculations used another rVV kernal table. Both the vdW kernal and rVV kernal tables were generated using \textcolor{red}{the} \textsc{Quantum-Espresso} package\cite{Giannozzi09,Pwscf14}. Based upon agreement with experimental formation enthalpies for Li$_2$C$_2$, LiC$_6$, and LiC$_{12}$, the optimized vdW density functional (optB88-vdW) \cite{Klimes10} method was used to calculated the final formation enthalpies for the low-enthalpy structures obtained in this work. We did not find any evidence that the vdW calculation provides improvement for the electronic band structures of Li-C compounds, although it \textcolor{red}{affects} the energy and force. Therefore, we used PBE for our density of states calculations. To compute phonon frequencies, we \textcolor{red}{also} used PBE.

\begin{widetext}

\begin{table}[h]
\caption{Formation energies (meV/atom) of lithium carbides. Both theoretical calculations (DFT and vdW \cite{Dion04,Grimme06,Thonhauser07,Romp09,Grimme10,Lee10,Klimes10,Klimes11,Sabatini12}) and experiments are at 1 atm pressure. In these calculations, the k-point meshes for Li$_2$C$_2$ (8 atoms cell), LiC$_6$ (7), and LiC$_{12}$ (13) are $6\times 6\times 8$, $8\times 8\times 8$ and $8\times 8\times 4$, respectively.
\label{tab:FEvdW}}
\begin{tabular}{@{}lcccccccccc@{}}
\hline
             & DFT-PBE & DFT-LDA & DFT-D2 & vdW  & vdw2 & rVV10 & optB88-vdW & optB86-vdW  & rev-vdW2 & Exp. \\
\hline
 Li$_2$C$_2$ &   4.3   &  52.1   &  -127  & -191 & -307 & -95.8 &   -127     &    -43.7    &  -48.0   & {-113}--{-177}\cite{Sangster07} \\
 LiC$_6$     &  -6.4   &  -61.0  & -96.2  & -7.7 & -6.2 & -10.9 &   -30.1    &    -28.6    &  -27.9   & -22.3\cite{Ohzuku93} \\
 LiC$_{12}$  &  -7.0   &  -37.5  & -59.1  & -8.9 & -8.5 & -10.4 &   -22.4    &    -21.9    &  -21.6   & -17.5\cite{Ohzuku93} \\
\hline
\end{tabular}
\end{table}
\end{widetext}

Our searches for the GS crystal structures of lithium carbides were performed at 0 GPa and a high pressure of 40 GPa. The zero temperature GS crystal of pure carbon is graphite at 0 GPa and is diamond at 40 GPa (see Fig. S1a in the supporting information), consistent with previous DFT calculations \cite{Wang10a,Zhu11}. The zero-temperature GS crystal structure of pure lithium is cI16 at 40 GPa (see Fig. S1b in the supporting information), which is determined in this work after comparing the enthalpies of six phases of pure lithium \cite{Hanfland00,Guillaume11,Schaeffer15}. At 0 GPa, however, the static enthalpy differences among FCC, 9R and HCP are too small (within 0.2 meV/atom) to determine which is most stable. Experiments \cite{Hanfland00,Guillaume11,Schaeffer15} suggested 9R would be the stable structure of lithium at 0 GPa. We chose FCC as the ground state at 0 GPa because it gave the lowest static enthalpy based on our opt88-vdW calculations. Zero point energies and temperature effects were not included in the enthalpies of pure elements as they were not included in the enthalpies of lithium carbides. The enthalpies of these GS structures of pure elements were used to calculate the formation enthalpies of predicted lithium carbides by Eq. \ref{eq:DeltaH} at the corresponding pressures. 

\begin{figure}[ht]
\centering
\includegraphics[width=3.2in]{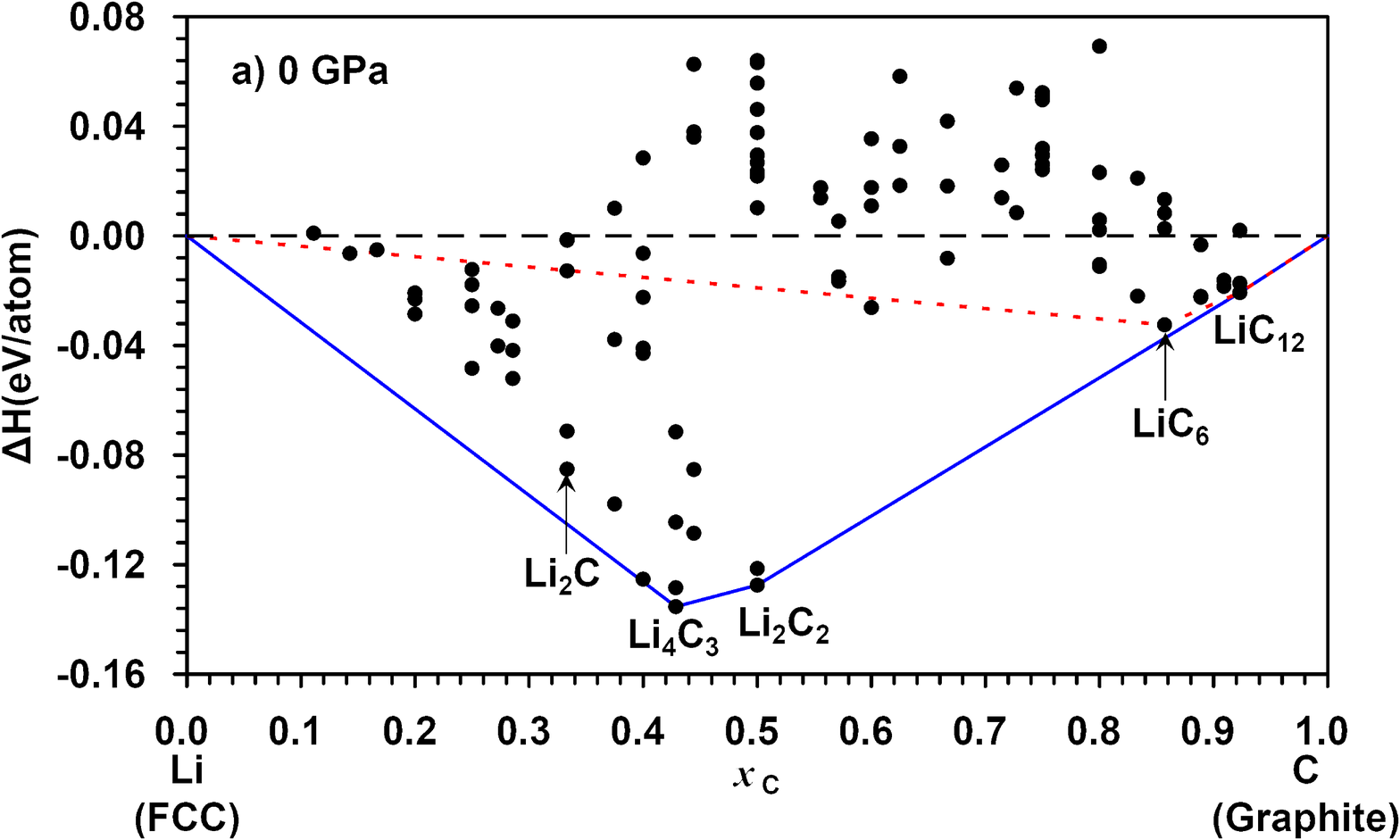}
\includegraphics[width=3.2in]{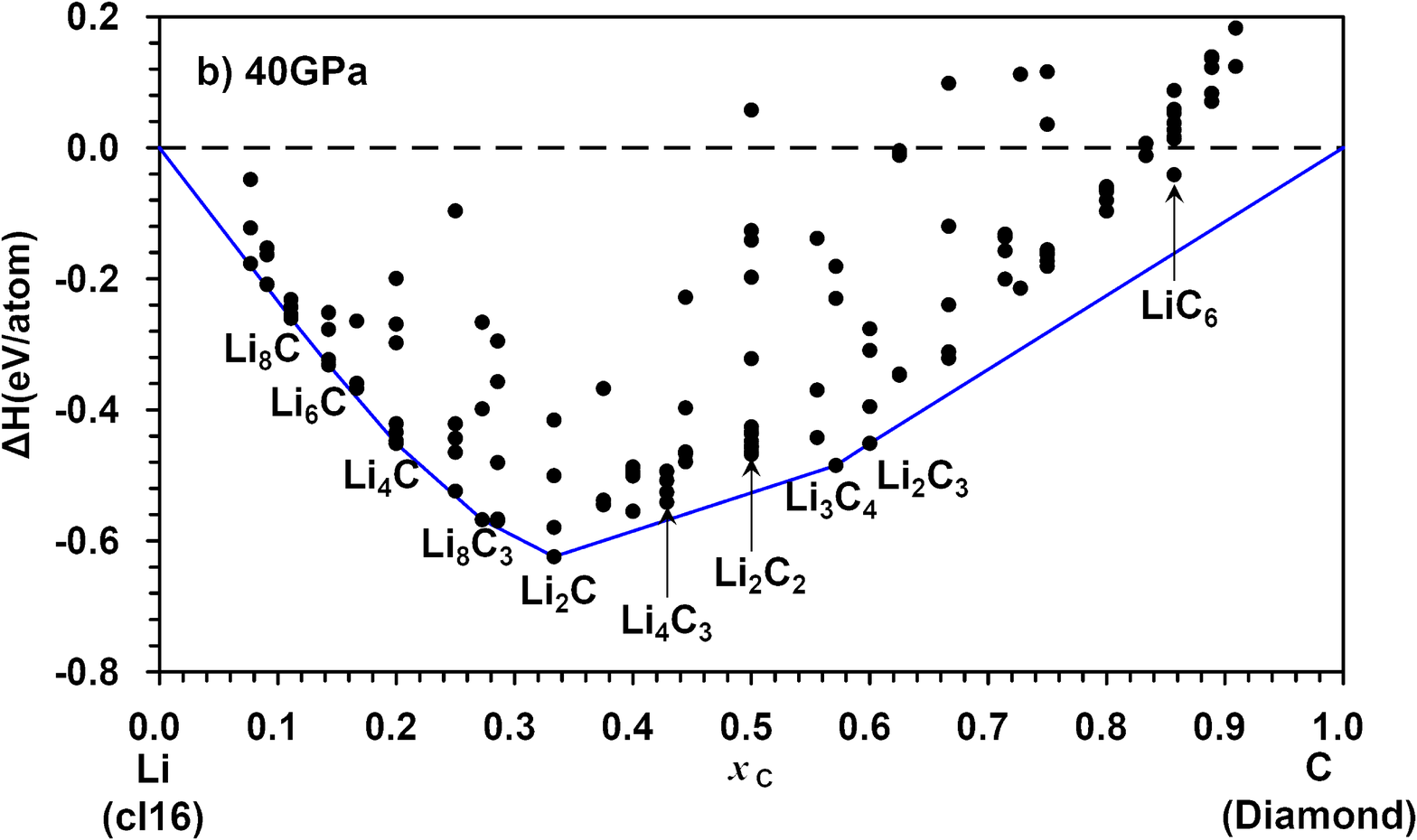}
\caption{Convex hulls of lithium carbides at pressures of a) 0 and b) 40 GPa. The solid circles indicate different structures. Those located on the convex hulls are thermodynamically stable at the corresponding pressures. The dashed lines in a) indicate a convex hull if only lithium graphite intercalation compounds were considered. The enthalpies of structures were calculated using the optB88-vdW method.}
\label{fig:Fig01}
\end{figure}

At 0 GPa, three thermodynamically stable lithium carbide phases (with stoichiometries Li$_4$C$_3$, Li$_2$C$_2$ and LiC$_{12}$) are identified according to the convex hull in Fig. \ref{fig:Fig01}a). The GS crystal structure of Li$_4$C$_3$ is monoclinic (symmetry \textit{C}2/\textit{m} in Hermann-Mauguin notation) with 14 atoms in its unit cell. This compound was reported previously through the lithiation of propyne with n-butyllithium in hexane\cite{West65,West69}. Later Jemmis et al.\cite{Jemmis77} proposed some interesting structures of Li$_4$C$_3$ molecular clusters through ab initio molecular orbital calculations. Carbon atoms exist as allylenide trimers in both our predicted Li$_4$C$_3$ crystal structures as well as the previously calculated molecular structures, however, the positions of lithium ions are different in the crystal and molecular forms. In crystalline Li$_4$C$_3$, there are 8 lithium atoms surrounding each carbon trimer within a Li-C distance of 2.20 {\AA}, and some of the lithium ions are shared by two or three carbon trimers, which was not considered \cite{Jemmis77} in the molecular structures of Li$_4$C$_3$. More crystallographic details of the \textit{C}2/\textit{m} Li$_4$C$_3$ can be found in Table S1 in the supporting information.

The crystal structure of the thermodynamically stable phase of Li$_2$C$_2$ obtained in this work is the same as the experimental structure determined by Ruschewitz and P\"ottgen \cite{Ruschewitz99}. It has a body-centered orthorhombic unit cell (\textit{Immm}) with 8 atoms. The crystal structure of the thermodynamically stable phase of LiC$_{12}$ takes the form of a lithium-intercalated graphite (LIG)\cite{Holzwarth83} structure, as expected. It has an AA$\alpha$ stacking sequence, the same as the structure proposed by some previous studies \cite{Holzwarth83,WangX12,Hazrati14}, where capital A indicates a layer of carbon (graphene) and the Greek letter $\alpha$ indicates a layer of lithium atoms. We found another metastable phase of LiC$_{12}$, which has a static formation enthalpy only 4 meV/atom higher than the thermodynamically stable one. The metastable LiC$_{12}$ is also a LIG structure with \textit{Immm} symmetry, but possesses a stacking sequence of A$\alpha$A$\beta$ (\textit{i.e.}, lithium atoms occupy two adjacent graphene layers).

Although LiC$_6$ can be easily synthesized experimentally at ambient conditions\cite{Drue13}, we find that this composition is not thermodynamically stable. The lowest-enthalpy form of LiC$_6$ (\textit{P}6/\textit{mmm}) is also a LIG structure with a stacking sequence of A$\alpha$, and is identical to the LiC$_6$ structure proposed previously \cite{WangX12,Hazrati14}. LiC$_6$ is metastable and decomposes into Li$_2$C$_2$ and LiC$_{12}$ according to the convex hull diagram shown in Fig \ref{fig:Fig01} a. Since the LIG structure of LiC$_6$ is very different from that of Li$_2$C$_2$ (carbon dimers), high energy barriers may exist in the pathways of LiC$_6$ decomposition\textcolor{red}{\cite{Konar15}} and can potentially explain the metastable observation of LiC$_6$ in experiments. Indeed, lattice dynamics calculations reveal that LiC$_6$ is mechanically stable at ambient pressure. We note that the GS structures of LiC$_4$, LiC$_5$, LiC$_6$, LiC$_8$, LiC$_{10}$ and LiC$_{12}$ at 0 GPa are all in LIG structures. If one only considers the carbon-rich side of the convex hull for LIG structures, \textit{i.e.}, above $x_{\text{C}}=0.75$ (see dashed lines in Fig. \ref{fig:Fig01} a), both LiC$_6$ and LiC$_{12}$ are ``thermodynamically stable''. LiC$_4$ and LiC$_5$ tend to decompose to LiC$_6$ plus pure lithium, and LiC$_8$ and LiC$_{10}$ tend to decompose to LiC$_6$ plus LiC$_{12}$, while LiC$_6$ would not decompose. Thus, in the absence of kinetic accessibility to the true GS, LiC$_6$ can be regarded as stable (in the family of LIGs), and also helps to explain why LiC$_6$ can be synthesized easily in experiments. It is possible that some other LIGs with higher carbon concentrations would be also thermodynamically stable at 0 GPa, as suggested by Hazrati et al.\cite{Hazrati14}, but those large structures exceed the atom limit range of this study. 

Structural diversity blossoms with increasing pressure, and at 40 GPa, we find seven thermodynamically stable lithium carbides with stoichiometries of Li$_8$C, Li$_6$C, Li$_4$C, Li$_8$C$_3$, Li$_2$C, Li$_3$C$_4$ and Li$_2$C$_3$, according to the convex hull in Fig. \ref{fig:Fig01}b. The GS crystal structures of Li$_6$C, Li$_4$C, Li$_8$C$_3$, Li$_2$C, Li$_3$C$_4$ and Li$_2$C$_3$ are in rhombohedral (\textit{R}-3\textit{m}),  tetragonal (\textit{I}4/\textit{m}), rhombohedral (\textit{R}-3\textit{m}), base-centered orthorhombic (\textit{Cmca}), body-centered orthorhombic (\textit{Immm}), and base-centered orthorhombic (\textit{Cmcm}) structures, respectively. Different from these relatively high-symmetry structures, the GS crystal structure of Li$_8$C is triclinic (\textit{P}-1) with 36 atoms in the unit cell. This stable large cell of Li$_8$C was derived from a dynamically unstable (with negative phonon frequencies) \textit{Cm} structure.

Except for these thermodynamically stable phases, some metastable crystal structures of lithium carbides are found at 40 GPa as well, \textit{i.e.}, a body-centered orthorhombic (\textit{Imm}2) and a base-centered orthorhombic (\textit{Cmcm}) structures of Li$_2$C$_2$ and a primitive orthorhombic (\textit{Pmmn}) structure of LiC$_6$. The body-centered orthorhombic structure (\textit{Imm}2) of Li$_2$C$_2$ has 24 atoms in the unit cell. It has a lower formation enthalpy than the previously reported Li$_2$C$_2$ structures \cite{Chen10,Benson13} at 40 GPa. The primitive cell orthorhombic structure of LiC$_6$ has \textit{Pmmn} symmetry with 14 atoms in the unit cell. The carbon atoms in \textit{Pmmn} LiC$_6$ exist as a mixed \textit{sp}$^2$-\textit{sp}$^3$ carbon framework. The crystallographic details of all the above structures can be found in Table S1 in the supporting information and detailed descriptions of these phases are given in the following sections. All of the newly predicted lithium carbides listed in Table S1 in the supporting information are dynamically stable from our DFPT calculations.

From the convex hulls of lithium carbides in Fig. \ref{fig:Fig01}, some exothermic chemical reactions among the thermodynamically stable phases are suggested. For example, the following reactions may happen below 40 GPa:

\begin{equation}
5\text{Li}_8\text{C}_3 + 2\text{Li}_3\text{C}_4 \xrightarrow{\text{40GPa,-2.82eV}} 23\text{Li}_2\text{C}
\label{eq:Li2C}
\end{equation}

\begin{equation}
\text{Li}_4\text{C} + 2\text{Li}_2\text{C} \xrightarrow{\text{40GPa,-0.38eV}} \text{Li}_8\text{C}_3
\label{eq:Li8C3}
\end{equation}

\subsection{Diverse Carbon Structures}

Throughout the thermodynamically stable phases (and some metastable ones) of lithium carbides, we found extreme diversity in carbon bonding with forms including carbon monomers, dimers, trimers, nanoribbons, sheets and frameworks. Carbon atoms within these different structures have very different electronic properties, and it may be possible to obtain novel pure carbon allotropes from these lithium carbides by removing all lithium atoms, particularly for the Li-C frameworks.

\begin{figure}[ht]
\centering
\includegraphics[width=3.2in]{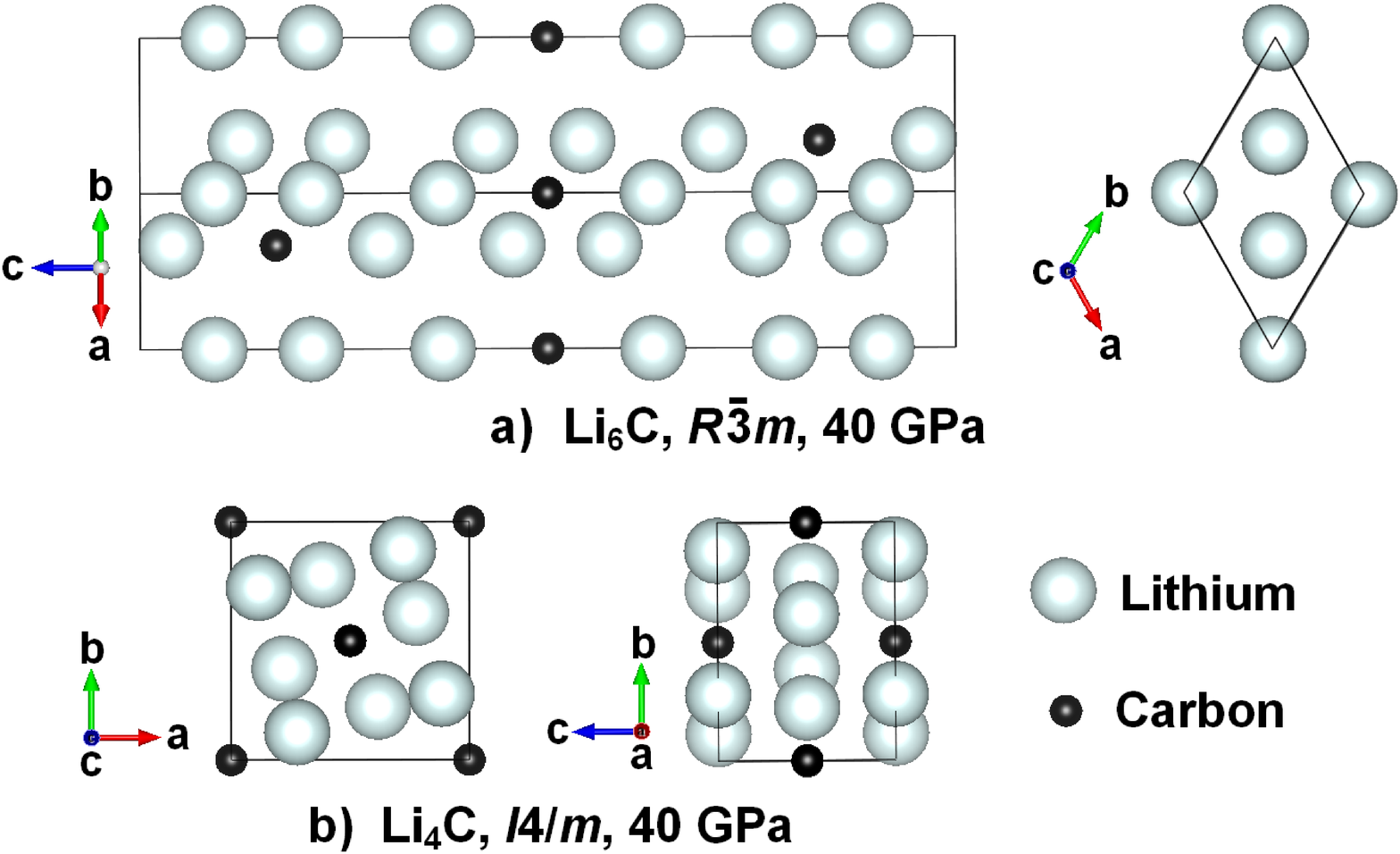}
\caption{The crystal structures of lithium carbides with carbon monomers. The image representations of lithium and carbon atoms are the same in the next crystal structure figures.}
\label{fig:Fig02}
\end{figure}

Carbon monomers (\textit{i.e.} methanide structures with no bonds to other carbon atoms) can be found in some thermodynamically stable phases at 40 GPa including Li$_8$C (\textit{P}-1), Li$_6$C (\textit{R}-3\textit{m}), Li$_4$C (\textit{I}4/\textit{m}) and Li$_8$C$_3$ (\textit{R}-3\textit{m}) (see in Fig. \ref{fig:Fig02}, with Li$_6$C and Li$_4$C as examples). Thermodynamically stable lithium carbides with carbon monomers are not found at 0 GPa and at 40 GPa they only exist within lithium-rich phases. All the lowest-enthalpy forms of Li$_8$C, Li$_6$C and Li$_4$C at 0 GPa have carbon dimers in their crystal structures and they are not thermodynamically stable. Nevertheless, if the lithium carbides with carbon monomers were synthesized at high pressures, they could possibly be quenched to ambient conditions, as is the case for Mg$_2$C \cite{Kurakevych13a,Kurakevych14} and Ca$_2$C \cite{Li15}. The minority phase synthesis of Li$_4$C was actually reported 40 years ago \cite{Chung72,Shimp73} through reaction between lithium and carbon vapor, but minimal yields have precluded definitive characterization \cite{Ruschewitz03,Kurakevych13a}.

The effective charges of carbon monomers in these thermodynamically stable lithium carbides vary from -1.99 to -2.13 for L\"owdin charges and from -2.80 to -3.12 for Bader charges and are very similar across the different methanide phases. Li$_4$C (\textit{I}4/\textit{m}) is metallic since it has a finite density of state (DOS) at the Fermi energy (FE) level (0.33 states/eV/fu or 0.22 states/eV/fu from PBE or GW calculations \cite{Caruso12}, respectively), which is different from other high-pressure thermodynamically stable methanides, e.g. Mg$_2$C (band gap 0.67 eV) \cite{Kurakevych13a,Kurakevych14} and Ca$_2$C (band gap 0.64 eV) \cite{Li15}. Li$_8$C (\textit{P}-1), Li$_6$C (\textit{R}-3\textit{m}), and Li$_8$C$_3$ (\textit{R}-3\textit{m}) are also metallic from our PBE calculations.

\begin{figure}[ht]
\centering
\includegraphics[width=3.2in]{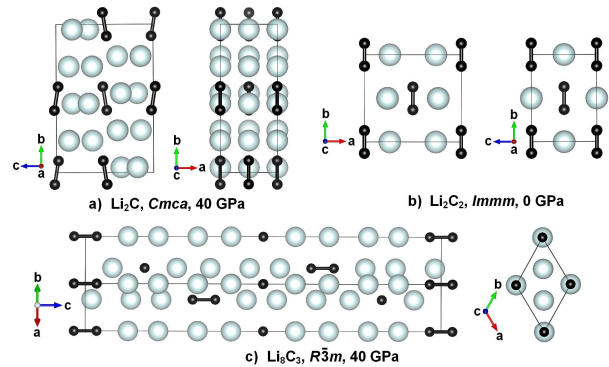}
\caption{The crystal structures of lithium carbides with carbon dimers. The line between two carbon atoms indicate a bond.}
\label{fig:Fig03}
\end{figure}

Carbon dimers exist in two thermodynamically stable phases, Li$_2$C (\textit{Cmca}) at 40 GPa and Li$_2$C$_2$ (\textit{Immm}) at 0 GPa (Fig. \ref{fig:Fig03} a and b). The bond lengths between two carbons in Li$_2$C$_2$ (\textit{Immm}) are 1.258{\AA} at 0 GPa and 1.237{\AA} at 40 GPa, while those in Li$_2$C (\textit{Cmca}) are 1.370{\AA} at 0 GPa and 1.373{\AA} at 40 GPa. The C-C bond length of dimeric C$_2$ unit varies from 1.19 to 1.48{\AA} in binary and ternary metal carbides \cite{Li89,Jeitschko89}. Bond lengths and bond types are mainly determined by the number of electrons transferred from metal to carbon atoms. Recalling that the covalent bond lengths between two carbons at ambient conditions are about 1.20{\AA} for triple bonds (C$\equiv$C), 1.33{\AA} for double bonds (C=C), and 1.54{\AA} for single bonds (C-C), we find that the carbon dimers in Li$_2$C$_2$ (\textit{Immm}) are ionic triple bonds (acetylide ion [C$\equiv$C]$^{2-}$) and those in Li$_2$C are ionic double bounds (ethenide ion [C=C]$^{4-}$). The DFT charges of carbon dimers in Li$_2$C$_2$ (\textit{Immm}) (L\"owdin -1.26 and Bader -1.74) and Li$_2$C (\textit{Cmca}) (L\"owdin -2.36 and Bader -3.12) are consistent with this. The calculated L\"owdin and Bader charges are always smaller than formal charge assignments.

Li$_2$C$_2$ (\textit{Immm}) is an insulator with a band gap of 3.3 eV or 6.4 eV from PBE or GW calculations, respectively, however Li$_2$C (\textit{Cmca}) is metallic with DOS of 0.36 states/eV/fu or 0.32 states/eV/fu at the FE level from PBE or GW calculations, respectively. So the carbon dimers in the insulating Li$_2$C$_2$ (\textit{Immm}) structure are indeed [C$\equiv$C]$^{2-}$ while the formal charge of [C=C]$^{4-}$ within metallic Li$_2$C (\textit{Cmca}) does not strictly apply. It is interesting to see that carbon monomers and carbon dimers coexist in the thermodynamically stable phase of Li$_8$C$_3$ (\textit{R}-3\textit{m},40 GPa) (Fig. \ref{fig:Fig03} c). The carbon dimers in Li$_8$C$_3$ (\textit{R}-3\textit{m}) also exist as double bonds based on their bond distances and charges. The carbon dimers in Li$_8$C$_3$ (R-3m) have bond distances of 1.352 {\AA} at 0 GPa and 1.394 {\AA} at 40 GPa, and at 40 GPa their L\"owdin and Bader charges are -2.36 and -3.48, respectively.

Acetylenic carbon ions (C$_2^{2-}$) are common in many binary metal carbides (e.g. Na$_2$C$_2$, K$_2$C$_2$, MgC$_2$ and CaC$_2$, etc.) \cite{Li89,Ruschewitz03,Li15}, whereas double bonded carbon dimers are unusual in binary carbides. They are found in UC$_2$ \cite{Li89} and recently predicted in a metastable phase of CaC \cite{Li15}. Such bonds are common in rare earth carbide halides (Y$_2$C$_2$Br$_2$, Y$_2$C$_2$I$_2$ and La$_2$C$_2$Br$_2$) \cite{Simon91} and ternary metal carbides (CeCoC$_2$, DyCoC$_2$ and U$_2$Cr$_2$C$_5$, etc.) \cite{Li89,Jeitschko89}. Since double bonded carbon dimers were found in two thermodynamically stable crystals Li$_2$C (\textit{Cmca}) and Li$_8$C$_3$ (\textit{R}-3\textit{m}), we predict them to form in lithium carbides using high-pressure experiments.

\begin{figure}[ht]
\centering
\includegraphics[width=3.2in]{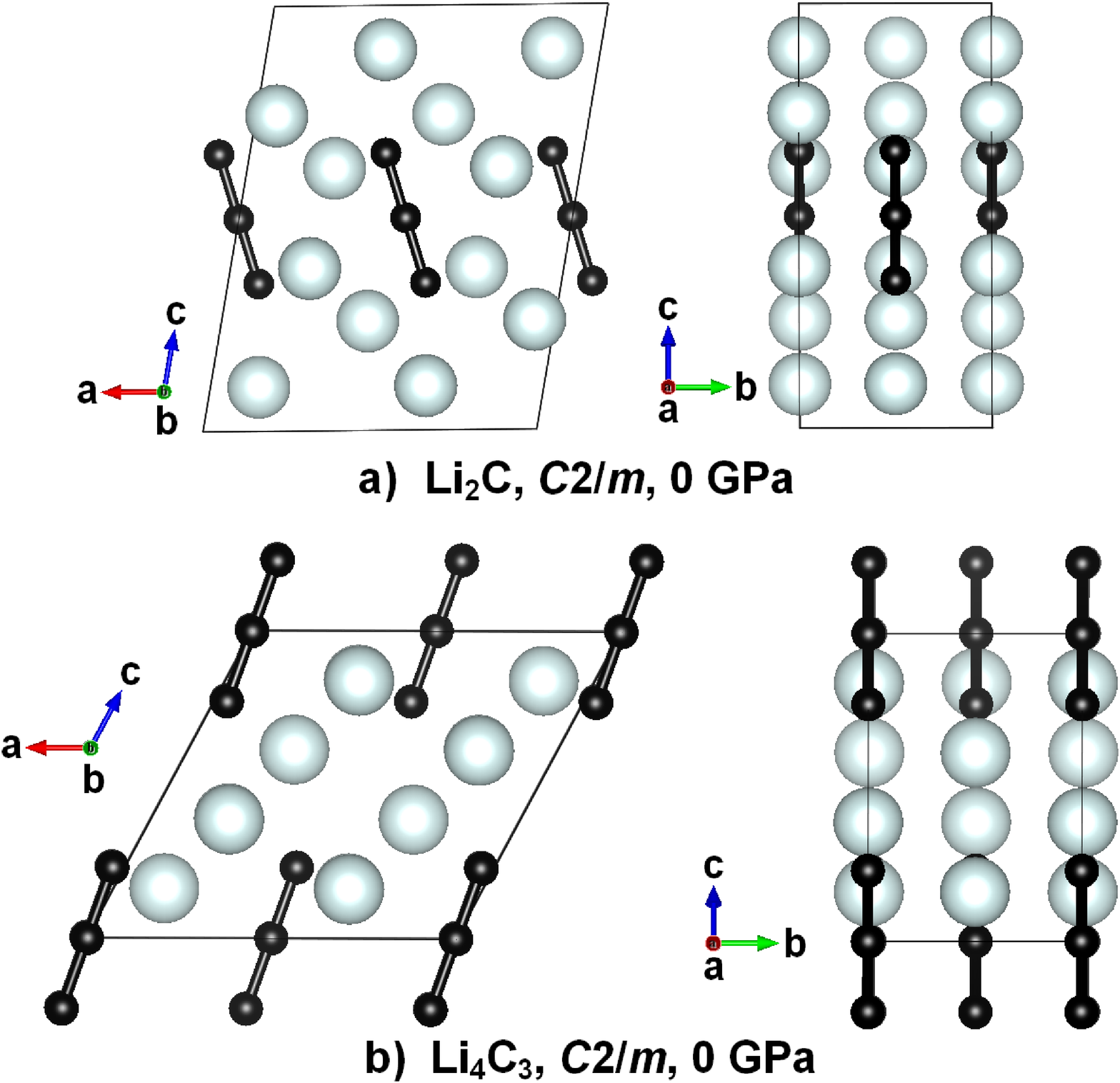}
\caption{The crystal structures of lithium carbides with carbon trimers.}
\label{fig:Fig04}
\end{figure}

Carbon trimers are found in a thermodynamically stable phase of Li$_4$C$_3$ (\textit{C}2/\textit{m}, 0 GPa) and a metastable phase of Li$_2$C (\textit{C}2/\textit{m}) (Fig. \ref{fig:Fig04}). We infer that the carbon trimers in both Li$_4$C$_3$ (\textit{C}2/\textit{m}) and Li$_2$C (\textit{C}2/\textit{m}) are allylenide-type [C=C=C]$^{4-}$ ions, since their bond lengths are around 1.34 {\AA}. This is confirmed by their L\"owdin and Bader charges. The L\"owdin and Bader charges on the carbon trimers of Li$_4$C$_3$ (\textit{C}2/\textit{m}) are -2.48 and -3.38 respectively and on those of Li$_2$C (\textit{C}2/\textit{m}) are -2.61 and -3.69 respectively. Li$_4$C$_3$ is a typical allylenide (C$_3^{4-}$) with a band gap of 0.98 eV or 2.2 eV from PBE or GW calculations, whereas Li$_2$C (\textit{C}2/\textit{m}) is metallic. With Mg$_2$C$_3$\cite{Strobel14} and Ca$_2$C$_3$\cite{Li15} included, we find three allylenides to be thermodynamically stable at ambient or high pressures. Although they are different in terms of chemistry, the crystal structure of stable Li$_4$C$_3$ has the same crystallographic symmetry (\textit{C}2/\textit{m}) as those of Mg$_2$C$_3$ and Ca$_2$C$_3$. The crystal structures of stable magnesium and calcium allylenides (isostructural in \textit{C}2/\textit{m}) have been confirmed recently by high-pressure experiments\cite{Strobel14,Li15}.

\begin{figure}[ht]
\centering
\includegraphics[width=3.2in]{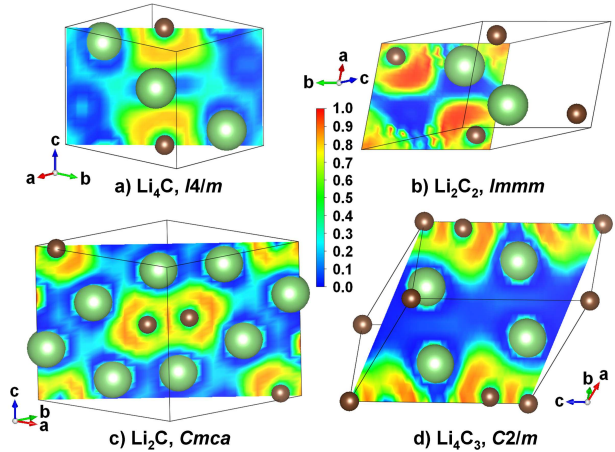}
\caption{Electron localization function (ELF) images in a) Li$_4$C (\textit{I}4/\textit{m}), b) Li$_2$C$_2$ (\textit{Immm}), c) Li$_2$C (\textit{Cmca}), and d) Li$_4$C$_3$ (\textit{C}2/\textit{m}). ELF = 1.0 corresponds to perfect localization and ELF = 0.0 corresponds to no electron. The smaller and the bigger spheres indicate carbon and lithium atoms respectively. The images were generated using \textsc{vesta} \cite{Momma11} based on DFT-PBE results.}
\label{fig:Fig05}
\end{figure}

Although Li$_4$C(\textit{I}4/\textit{m}), Li$_2$C(\textit{Cmca}), Li$_4$C$_3$(\textit{C}2/\textit{m}) and Li$_2$C$_2$(\textit{Immm}) are expected to be insulating based on their structures and formal charge balance rules, Li$_4$C(\textit{I}4/\textit{m}) and Li$_2$C(\textit{Cmca}) are actually metallic based on our theoretical calculations. To better understand this phenomenon, we compare the electron localization functions (ELFs) of Li$_4$C (\textit{I}4/\textit{m}), Li$_2$C(\textit{Cmca}), Li$_4$C$_3$(\textit{C}2/\textit{m}) and Li$_2$C$_2$(\textit{Immm}) in Fig. \ref{fig:Fig05}. For Li$_2$C$_2$, electrons are strongly localized to carbon orbitals, thus this structure is insulating. Electron localization remains high in Li$_4$C$_3$ (\textit{C}2/\textit{m}), but slightly smaller than in Li$_2$C$_2$ (\textit{Immm}), which is consistent with the semiconducting nature of Li$_4$C$_3$ (\textit{C}2/\textit{m}). In both Li$_4$C (\textit{I}4/\textit{m}) and Li$_2$C (\textit{Cmca}), electrons are far more delocalized than in Li$_2$C$_2$ (\textit{Immm}) and in Li$_4$C$_3$ (\textit{C}2/\textit{m}). This result helps to explain why Li$_4$C (\textit{I}4/\textit{m}) and Li$_2$C (\textit{Cmca}) are metallic.

\begin{figure}[h]
\centering
\includegraphics[width=3.2in]{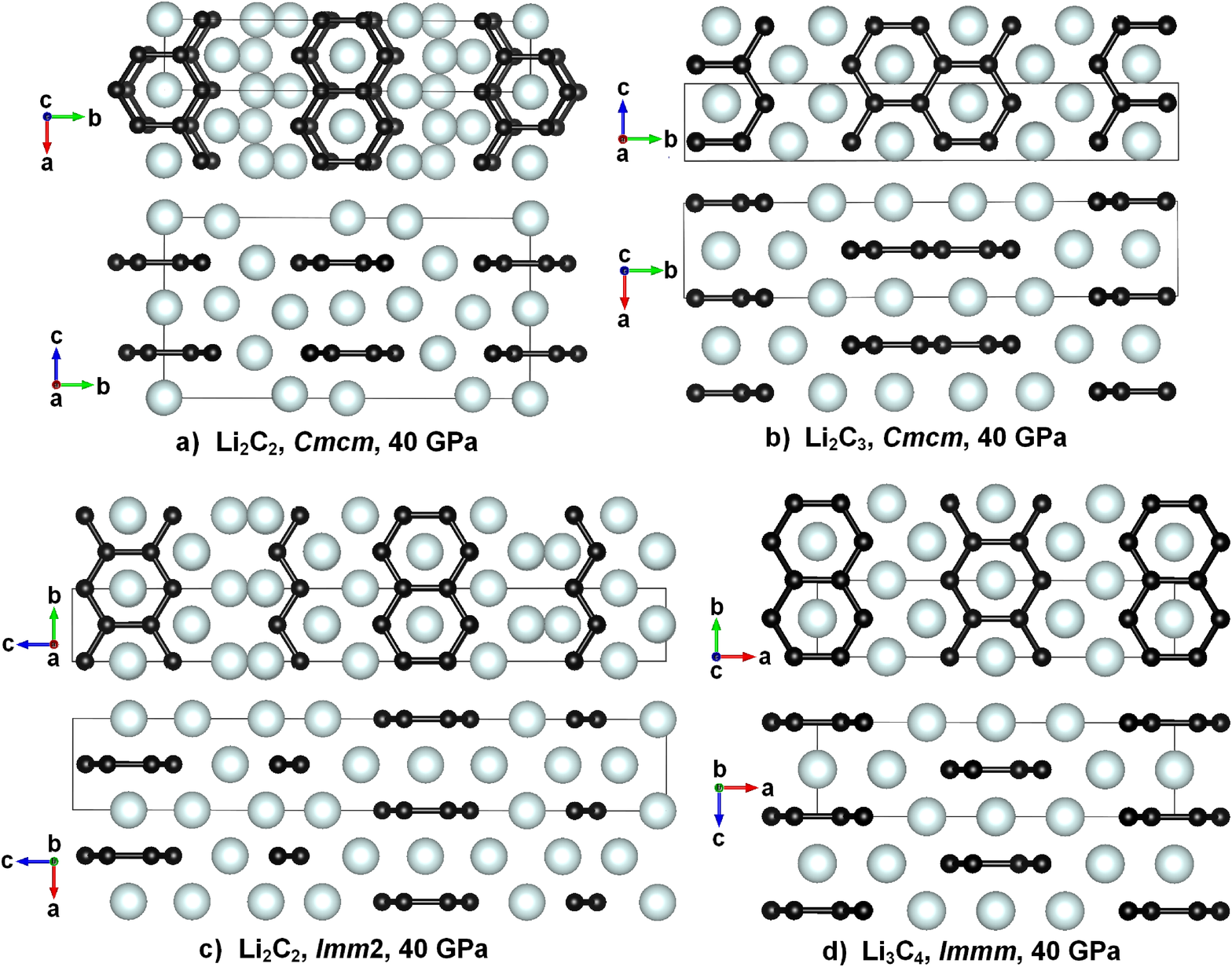}
\caption{The crystal structures of lithium carbides with carbon nanoribbons.}
\label{fig:Fig06}
\end{figure}

Carbon nanoribbons were found in two thermodynamically stable phases (Li$_3$C$_4$, \textit{Immm} and Li$_2$C$_3$, \textit{Cmcm} at 40 GPa) as well as some metastable phases (Li$_2$C$_2$, \textit{Imm}2 and \textit{Cmcm}) (Fig. \ref{fig:Fig06}). The carbon nanoribbons can be formed by one, two or three zig-zag carbon chains, and thus have different widths. There are two main types of carbon atoms in these carbon nanoribbons. The first type of carbon atoms lie on the sides of ribbons and only have two bonds to other carbon atoms, whereas the other type of carbon is in the middle of the ribbons, and has three bonds to adjacent carbon atoms. The side carbon atoms (L\"owdin {-0.74}--{-0.87} and Bader {-0.77}--{-1.24}) have more electron density than the middle ones (L\"owdin {-0.28}--{-0.36} and Bader {-0.40}--{-0.52}). At 40 GPa, the bond lengths between carbon atoms in all ribbons do not differ greatly. They range between {1.43}--{1.49} {\AA}. The ribbons must contain carbon atoms with exclusively \textit{sp}$^2$ hybridization since the atoms in each ribbon lie exactly within the same plane. The extra electrons go into the \textit{sp}$^2$ orbitals and enlarged the C-C bond lengths in comparison with the bonds of graphite (1.39 {\AA}) at the same pressure of 40 GPa.

\begin{figure}[ht]
\centering
\includegraphics[width=3.2in]{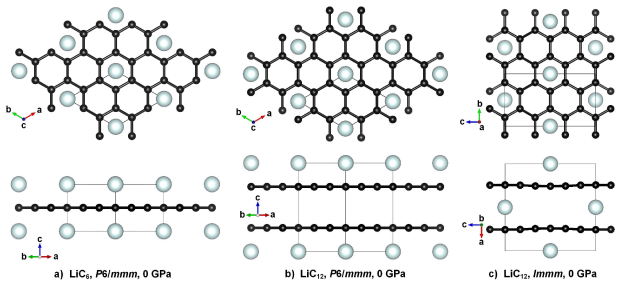}
\caption{The crystal structures of lithium carbides with carbon sheets.}
\label{fig:Fig07}
\end{figure}

Carbon sheets are found in graphite intercalation compounds (GICs) such as LiC$_6$ and LiC$_{12}$ (Fig. \ref{fig:Fig07}). At 0 GPa, the C-C bond lengths are {1.439}--{1.442} {\AA} in LiC$_6$ (\textit{P}6/\textit{mmm}) and {1.431}--{1.434} {\AA} in LiC$_{12}$ (\textit{P}6/\textit{mmm} and \textit{Immm}). These bond lengths indicate that carbon atoms in the carbon sheets are \textit{sp}$^2$ hybridized. Both L\"owdin and Bader charges show that carbon atoms in LiC$_6$ (\textit{P}6/\textit{mmm}, L\"owdin -0.11 and Bader -0.15) have more electrons than in LiC$_{12}$ (\textit{P}6/\textit{mmm} and \textit{Immm}, L\"owdin {-0.033}--{-0.035} and Bader {-0.070}--{-0.079}). Without lithium atoms, the C-C bond lengths would be 1.424 {\AA}, as in graphite at 0 GPa. These results show a typical tendency that with increasing lithium content between carbon sheets, more electrons are transferred to the \textit{sp}$^2$ orbitals, thus increasing the C-C bond lengths.

\begin{figure}[ht]
\centering
\includegraphics[width=3.2in]{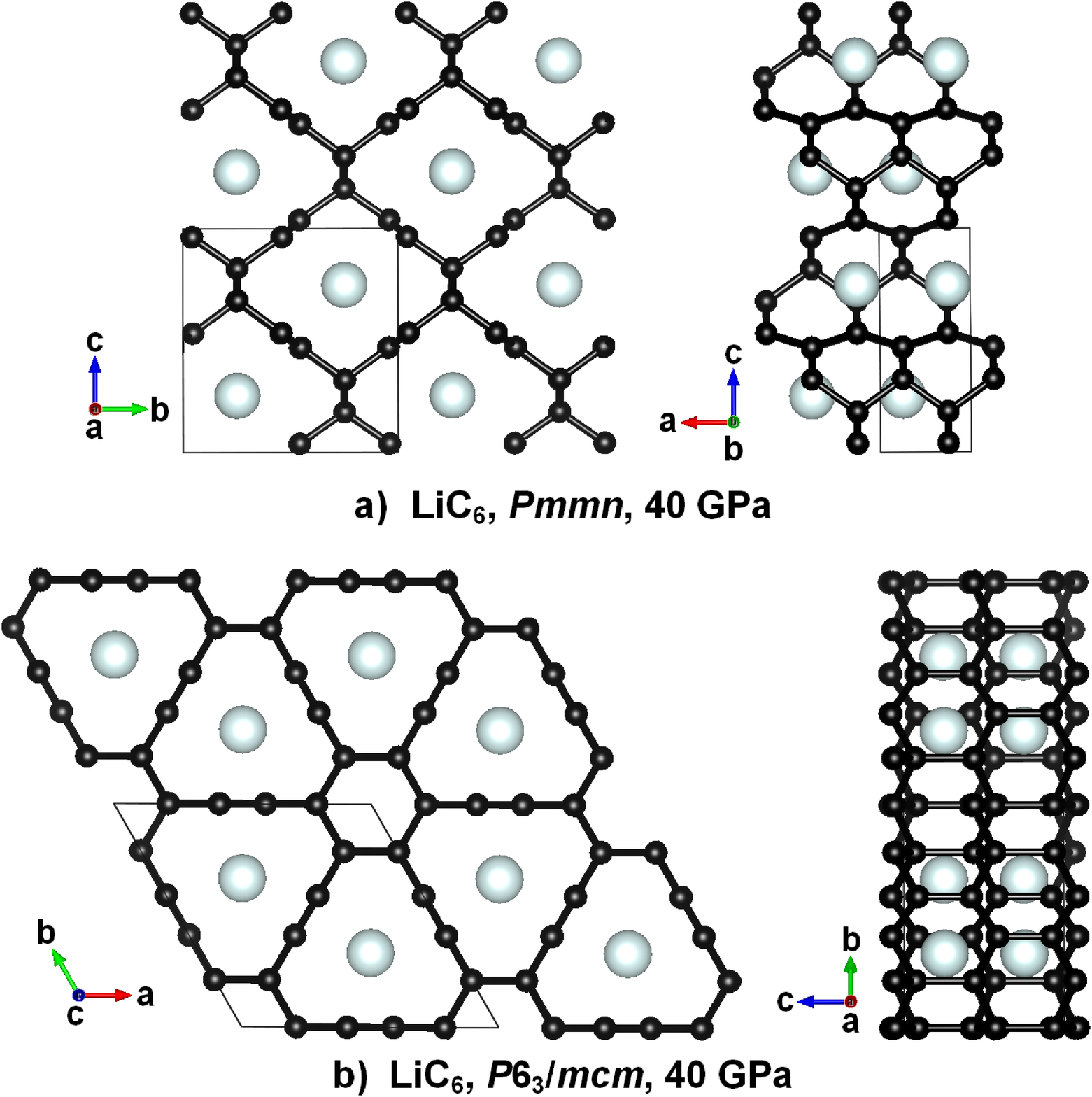}
\caption{The crystal structures of lithium carbides with carbon frameworks.}
\label{fig:Fig08}
\end{figure}

In this work, we also found that carbon frameworks exist in some low-enthalpy lithium carbides of LiC$_6$ (\textit{Pmmn} and \textit{P}6$_3$/\textit{mcm}) at 40 GPa (Fig. \ref{fig:Fig08}). Despite the fact that the metal valences are different in LiC$_6$ (\textit{Pmmn}) and a previously predicted metastable phase of CaC$_6$\cite{Li13}, these two compounds have identical crystal structures. Although these lithium carbides are not thermodynamically stable, they are metastable with all positive phonon frequencies. These carbon frameworks are formed by zig-zag carbon chains along a channel that contains lithium ions. We investigated some other lithium carbides with carbon framework structures in LiC$_5$, LiC$_6$, LiC$_8$ and LiC$_{10}$, but they have much higher formation enthalpies than the \textit{Pmmn} and \textit{P}6$_3$/\textit{mcm} phases of LiC$_6$. We will discuss the potential phase transitions from LiC$_6$ (\textit{P}6/\textit{mmm}) to LiC$_6$(\textit{Pmmn}) in the following section.

\subsection{Li$_2$C$_2$ and LiC$_6$}

Li$_2$C$_2$ and lithium-intercalated graphites (LIGs) are the most easily synthesized lithium carbides at ambient conditions \cite{Drue13}. Both Li$_2$C$_2$ and LIGs are unstable and tend to decompose to other compositions at high pressures according the convex hulls in Fig. \ref{fig:Fig01}. At 40 GPa, the most energetically favorable decomposition pathways for Li$_2$C$_2$ and LiC$_6$ (an example for LIGs) are as follows,

\begin{equation}
5\text{Li}_2\textcolor{red}{\text{C}}_2 \xrightarrow{\text{high pressure}} 2\text{Li}_2\text{C} + 2\text{Li}_3\text{C}_4
\label{eq:Li2C2}
\end{equation}

\begin{equation}
2\text{LiC}_6 \xrightarrow{\text{high pressure}} \text{Li}_2\text{C}_3 + 9\text{C}
\label{eq:LiC6}
\end{equation}

In addition to decomposition, the ambient-pressure ground states of Li$_2$C$_2$ (\textit{Immm}) and LiC$_6$ (\textit{P}6/\textit{mmm}) would also tend to transform to other isocompositional metastable phases at high pressures.

\begin{figure}[ht]
\centering
\includegraphics[width=3.2in]{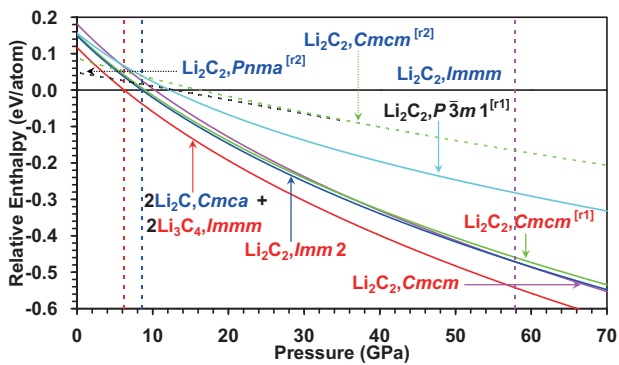}
\caption{Relative enthalpies of several phases of Li$_2$C$_2$ and its decomposition. \textcolor{red}{The dumbbell-containing and ribbon-containing structures are indicated by blue and red texts, respectively. The \textit{P}-3\textit{m}1$^{[r1]}$ is a sheet-containing structure. [r1] and [r2] indicate references \cite{Benson13} and \cite{Efthimiopoulos15}, respectively}. The enthalpies were calculated using the optB88-vdW method \cite{Klimes10}. In these calculations, the k-point meshes for Li$_2$C$_2$ (\textit{Immm}, 8 atoms cell), Li$_2$C (\textit{Cmca}, 24), Li$_3$C$_4$ (\textit{Immm}, 14), Li$_2$C$_2$ (\textit{Imm}2, 24), Li$_2$C$_2$ (\textit{Cmcm}, 32), Li$_2$C$_2$ (\textit{Cmcm}$^{[r1]}$, 8), Li$_2$C$_2$ (\textit{P}-3\textit{m}1$^{[r1]}$, 4), \textcolor{red}{Li$_2$C$_2$ (\textit{Pnma}$^{[r2]}$, 16) and Li$_2$C$_2$ (\textit{Cmcm}$^{[r2]}$, 8)} are $6\times 6\times 10$, $8\times 4\times 6$, $4\times 12\times 10$, $10\times 12\times 2$, $12\times 2\times 4$, $10\times 12\times 4$, $12\times 12\times 6$, \textcolor{red}{$6\times 6\times 6$ and $8\times 8\times 6$}, respectively. }. 
\label{fig:Fig09}
\end{figure}

Li$_2$C$_2$ (\textit{Immm}) would transform to two other phases (\textit{P}-3\textit{m}1 and \textit{Cmcm}) at high pressures according to previous studies \cite{Chen10,Benson13}. Since we find two more stable phases of Li$_2$C$_2$ (\textit{Imm}2 and \textit{Cmcm}) at 40 GPa, \textcolor{red}{these transition paths are more thermodynamically favorable}. Although the symmetries are the same, our \textit{Cmcm} structure of Li$_2$C$_2$ is different from that determined in Ref. \textcolor{red}{[52]}. Each carbon ribbon in our \textit{Cmcm} structure is formed by 2 zig-zag chains (Fig. \textcolor{red}{\ref{fig:Fig06}a}), while the previous determined \textit{Cmcm} structure contains just one zig-zag chain. Based on the static enthalpies calculated from vdW density functional (optB88-vdW) theory \cite{Klimes10}, Li$_2$C$_2$ (\textit{Immm}) would decompose to Li$_2$C (\textit{Cmca}) and Li$_3$C$_4$ at 6.2 GPa (Fig. \ref{fig:Fig09}). If decomposition under local equilibrium is avoided, Li$_2$C$_2$ (\textit{Immm}) would transform to our \textit{Imm}2 and \textit{Cmcm} structures at 8.6 GPa and 57.8 GPa, respectively. Carbon nanoribbons are energetically favorable over carbon dimers in Li$_2$C$_2$ at high pressures. \textcolor{red}{Recent experimental work on Li$_2$C$_2$ indicates a phase transition from \textit{Immm} to a dumbbell-containing \textit{Pnma} structure, and another dumbbell-type \textit{Cmcm} structure was predicted at higher pressure \cite{Efthimiopoulos15}. In view of this work, we also did calculations for two C$_2$ dumbbell-containing structures proposed by Efthimiopoulos et al. \cite{Efthimiopoulos15}. Our vdW calculations indicate both these \textit{Pnma} and \textit{Cmcm} structures are metastable(Fig. \ref{fig:Fig09}). Under hydrostatic compression experiments \cite{Nylen12,Efthimiopoulos15}, the ambient stable Li$_2$C$_2$ (\textit{Immm}) phase survived until 15 GPa, before transforming to \textit{Pnma} structure. While disproportion is more favorable based on our calculations, this transition does represent the most thermodynamically favorable pathway amongst the dumbbell-type Li$_2$C$_2$ structures, and can be understood by considering that room-temperature may not provide sufficient thermal energy to access the more energetically favorable Li$_2$C and Li$_3$C$_4$ phases. The experimentally observed amorphization transition above ~25 GPa \cite{Nylen12,Efthimiopoulos15} can also be understood in this way and likely represents a frustrated transition that is kinetically hindered at room temperature.}

\begin{figure}[ht]
\centering
\includegraphics[width=3.2in]{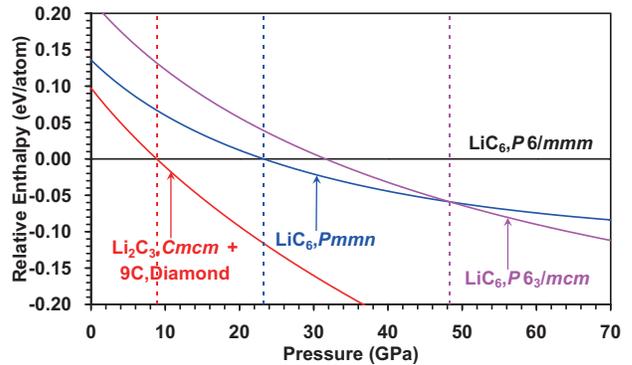}
\caption{Relative enthalpies of several phases of LiC$_6$(\textit{P}6/\textit{mmm}, \textit{Pmmn} and \textit{P}6$_3$/\textit{mcm}) and for decomposition to Li$_2$C$_3$ (\textit{Cmcm}) and carbon diamond. The enthalpies were calculated using the optB88-vdW method \cite{Klimes10}. In these calculations, the k-point meshes for LiC$_6$ (\textit{P}6/\textit{mmm}, 7 atoms cell), Li$_2$C$_3$ (\textit{Cmcm}, 20), Diamond C (2), LiC$_6$ (\textit{Pmmn}, 14), and LiC$_6$ (\textit{P}6$_3$/\textit{mcm}, 28) are $8\times 8\times 10$, $10\times 2\times 12$, $14\times 14\times 14$, $12\times 6\times 6$, and $6\times 6\times 6$, respectively.}
\label{fig:Fig10}
\end{figure}

According to our computed formation enthalpies, LiC$_6$ (\textit{P}6/\textit{mmm}) transforms to the \textit{Pmmn} and \textit{P}6$_3$/\textit{mcm} structures, at pressures of 23.3 GPa and 48.4 GPa, respectively (Fig. \ref{fig:Fig10}). Similar to the situation in Li$_2$C$_2$, the decomposition of LiC$_6$ (\textit{P}6/\textit{mmm}) to Li$_2$C$_3$ (\textit{Cmcm}) and C (diamond) is predicted at 8.8 GPa, \textcolor{red}{and a} phase transition of LiC$_6$ from \textit{P}6/\textit{mmm} to \textit{Pmmn} would occur if the barrier energy is higher in the pathway of decomposition than in the pathway of phase transition. In any case, the 3D carbon structures (diamond or framework in \textit{Pmmn} LiC$_6$) would be formed from the carbon sheets in LiC$_6$ at high pressures.

\section{Conclusions}

We predict the convex hulls of lithium carbides at 0 GPa and 40 GPa. Based on the convex hulls in the range from Li$_{12}$C to LiC$_{12}$, three phases (including Li$_4$C$_3$, Li$_2$C$_2$ and LiC$_{12}$) at 0 GPa and seven phases (including Li$_8$C, Li$_6$C, Li$_4$C, Li$_8$C$_3$, Li$_2$C, Li$_3$C$_4$, and Li$_2$C$_3$) at 40 GPa are identified as thermodynamically stable at the corresponding pressures. Although LiC$_6$ is not thermodynamically stable from our calculations, it is conditionally (within LIGs) stable at 0 GPa. Our results indicate all the stable phases at 0 GPa are metastable at 40 GPa and all the stable phases at 40 GPa are metastable at 0 GPa.

Carbon monomers exist in four high-pressure, thermodynamically stable phases of Li$_8$C (\textit{P}-1, 40 GPa), Li$_6$C (\textit{R}-3\textit{m}, 40 GPa), Li$_4$C(\textit{I}4/\textit{m}, 40 GPa) and Li$_8$C$_3$(\textit{R}-3\textit{m}, 40 GPa). Carbon dimers can be found in three thermodynamically stable phases of Li$_8$C$_3$ (\textit{R}-3\textit{m}, 40 GPa),Li$_2$C (\textit{Cmca}, 40 GPa) and Li$_2$C$_2$ (\textit{Immm}, 0 GPa). The carbon-carbon bonds in carbon dimers have either triple bonds (in Li$_2$C$_2$) or double bonds (Li$_8$C$_3$ and Li$_2$C). Li$_4$C$_3$ (\textit{C}2/\textit{m}) is predicted as a typical allylenide with carbon trimers ([C=C=C]$^{4-}$) and it is thermodynamically stable at 0 GPa. Although Li$_4$C (\textit{I}4/\textit{m}) and Li$_2$C (\textit{Cmca}) are expected to be insulating based on formal charge balance rules, they are metallic even at 0 GPa, which is different from Li$_4$C$_3$ (\textit{C}2/\textit{m}) and Li$_2$C$_2$ (\textit{Immm}). The band gaps of Li$_4$C$_3$ (\textit{C}2/\textit{m}) and Li$_2$C$_2$ (\textit{Immm}) are 2.2 eV and 6.4 eV, respectively.

Carbon nanoribbons are frequently found in the high-pressure thermodynamically stable and metastable phases of lithium carbides with moderate carbon concentrations (Li$_2$C$_2$, Li$_3$C$_4$, and Li$_2$C$_3$). Carbon ribbons may exist with different widths. We predict all the phases with carbon ribbons to be metallic.

Carbon sheets are the fundamental carbon structures within graphite intercalation compounds. All of the carbon atoms within ribbons and sheets maintain \textit{sp}$^2$ hybridization. With increasing lithium \textcolor{red}{content} in LIGs, more electrons are transferred into the \textit{sp}$^2$ orbitals of carbon atoms, which makes the carbon-carbon bond \textcolor{red}{lengths} longer. We find that carbon frameworks only prefer to exist at the composition LiC$_6$. A typical tendency is that the dimensionality of the stable carbon form existing in each structure increases with the increasing of carbon concentration at the same pressure. At 0 GPa, the dimensionality of carbon increases from 1D (dimers and trimers) to 2D (sheets), while at 40 GPa, carbon dimensionality increases from 0D (monomer), 1D(dimer), 2D(ribbons) to 3D (frameworks or \textit{sp}$^3$ structures as in diamond). 

Pressure is a crucial variable for the GS crystal structures of lithium carbides and reveals a dramatic range of chemical diversity. Carbon ribbons can be obtained by compressing Li$_2$C$_2$ to larger than 6.2 GPa and 3D carbon structures (frameworks or diamond) may be obtained by compressing LiC$_6$ to larger than 8.8 GPa. If chemical disproportionation occurs, the pressure needed for structure transformations in Li$_2$C$_2$ and LiC$_6$ would be much lower. We expect that these predictions will inspire experimental efforts.

\section{Acknowledgments}
\label{acknowledgments}
This work is supported by DARPA under grant No. W31P4Q1310005. Most of our DFT computations were performed on the supercomputer Copper of DoD HPCMP Open Research Systems under project No. ACOMM35963RC1. REC is supported by the Carnegie Institution for Science and by the European Research Council Advanced Grant ToMCaT. We thank Dr. R. Hoffmann at Cornell University, Dr. P. Ganesh at Oak Ridge National Laboratory, and Dr. L. Shulenburger at Sandia National Laboratories for useful discussions and comments on this manuscript.

\clearpage

\end{document}